\documentclass[%
 reprint,
 amsmath,amssymb,
 aps,
]{revtex4-1}

\usepackage{graphicx}% Include figure files
\usepackage{dcolumn}% Align table columns on decimal point
\usepackage{bm}% bold math

\usepackage[linktocpage=true,
  colorlinks=true, 
  pdfborder={0 0 0},
  linkcolor=blue,
  citecolor=red,
  filecolor=yellow,
  urlcolor=blue,
  bookmarks,
  pdfauthor={},
]{hyperref}

\begin{document}

\preprint{APS/123-QED}

\title{Comparison of Green's functions for transition metal atoms using \\ self-energy functional theory and coupled-cluster singles and doubles (CCSD)}

\author{Taichi Kosugi}
\author{Hirofumi Nishi}
\author{Yoritaka Furukawa}
\author{Yu-ichiro Matsushita}
\affiliation{
Department of Applied Physics, The University of Tokyo, Tokyo 113-8656, Japan
}

\date{\today}% It is always \today, today,

\begin{abstract}
We demonstrate in the present study that self-consistent calculations based on the self-energy functional theory (SFT) are possible for the electronic structure of realistic systems in the context of quantum chemistry.
We describe the procedure of a self-consistent SFT calculation in detail and perform the calculations for isolated $3 d$ transition metal atoms from V to Cu as a preliminary study.
We compare the one-particle Green's functions (GFs) obtained in this way and those obtained from the coupled-cluster singles and doubles (CCSD) method.
Although the SFT calculation starts from the spin-unpolarized Hartree--Fock (HF) state for each of the target systems,
the self-consistency loop correctly leads to degenerate spin-polarized ground states.
We examine the spectral functions in detail to find their commonalities and differences among the atoms by paying attention to the characteristics of the two approaches.
It is demonstrated via the two approaches that calculations based on the density functional theory (DFT) can fail in predicting the orbital energy spectra for spherically symmetric systems.
It is found that the two methods are quite reliable and useful beyond DFT.
\end{abstract}

\maketitle

\section{\label{sec:introduction} Introduction}

Electronic-structure calculations based on the density functional theory (DFT) have been successful for  quantitatively accurate predictions and explanations of the properties of various electronic systems.
The properties of interacting electrons are analyzed usually by mapping the problem to another one where non-interacting electrons feel a mean-field-like potential,
leading to the Kohn-Sham (KS) equation\cite{bib:76,bib:77}.
DFT-based calculations have, however, a tendency to fail in describing the material properties even qualitatively for strongly correlated systems.
To remedy such drawbacks, various approaches have been proposed.
The dynamical mean-field theory (DMFT)\cite{PhysRevB.45.6479} is one of the most powerful alternatives allowing for reliable description of electronic correlation.
It was proposed originally for solving the Hubbard model by mapping it to the Anderson model with an effective medium. 
DMFT has been, however, applied as well to electronic-structure calculations\cite{bib:3127} in conjunction with DFT.

Potthoff\cite{bib:3694,Potthoff_VCA} proposed an approach for description of correlated electronic systems at finite temperatures.
It treats the self-energy of a target many-electron system as a fundamental quantity from which the various physical quantities are calculated.
It is thus called the self-energy functional theory (SFT),
in contrast to DFT with the electron density as its fundamental quantity.
SFT approach has been applied to study the Hubbard models\cite{bib:3695,bib:3499,bib:3540,bib:3059} and their metal-insulator transition\cite{bib:3503,bib:3548} with introducing variational parameters.
SFT calculations for realistic systems\cite{bib:3506,bib:3505,bib:3502} have also been performed by using the parameters extracted from the results of DFT calculations.

Alongside the development of the DFT-based approaches,
many sophisticated approaches based on the wave function theory have been proposed for quantum chemistry calculations.
The coupled-cluster (CC) methods\cite{Helgaker} are widely accepted ones for achieving high accuracy and suppressing computational cost simultaneously depending on target correlated systems being considered.
Amongst them, the Green's functions (GFs) within the framework of the coupled-cluster singles and doubles (CCSD)\cite{Kowalski14,Kowalski16} method have been drawing attention recently.
We adopted this approach to calculate the CCSD GFs for light atoms\cite{GF_CCSD_and_FCI_for_light_atoms} and periodic systems\cite{GF_CCSD_for_periodic_systems}
and examined their features systematically.

The second-order GF theory (GF2)\cite{bib:4432,bib:3578} has been deserving attention recently in the quantum chemistry community.
It incorporates the self-energy diagrams self-consistently up to second order, that is,
the frequency-dependent second-order Coulomb and exchange diagrams in addition to the frequency-independent Hartree--Fock (HF) self-energy. 
GF2 has been applied not only to isolated systems\cite{bib:3581,bib:3704} but also to periodic systems.\cite{bib:3689}
In combination with GF2, the self-energy embedding theory (SEET)\cite{bib:3701,bib:3702,bib:3996} constructs the self-energy for the whole Hilbert space of a target system by embedding the exact ones for the correlated subspaces block diagonally,
to which those obtained via GF2 are added for the off-diagonal components.
SFT method can be regarded to be a kind of SEET method, as explained later.

Photoelectron spectroscopy, whose history dates back to the end of the 19th century\cite{photoelectron_Hertz,photoelectron_Hallwachs,photoelectron_Lenard}, has become an active research field today.
Experiments on the photoelectric effects make use of various kinds of measurements such as ultraviolet photoelectron spectroscopy (UPS), X-ray photoelectron spectroscopy (XPS), and angle-resolved photoemission spectroscopy (ARPES) for elucidating the properties of materials.
From a theoretical viewpoint, these effects are often explained and predicted based on the so-called sudden approximation\cite{bib:4070},
in which the observed spectra are given via the one-particle GF of a many-electron system.
The development of reliable calculation methods for GFs and comparison between them are thus important.

Although SFT approach have been applied to pure or DFT-based model systems, as stated above,
an electronic-structure calculation based purely on SFT has not been reported, to our best knowledge.
Therefore we demonstrate in the present study the applicability of SFT to realistic systems by describing the calculation procedure in detail.
We perform self-consistent SFT calculations for isolated transition metal atoms from V to Cu and examine the features of their calculated one-particle GFs by comparing those obtained via CCSD method.

This paper is organized as follows.
In Sect. \ref{sec:methods}, we explain the calculation methods adopted in the present study,
focusing particularly on those for the SFT calculations.
In Sect. \ref{sec:computational_details}, we describe the details of our computation.
In Sect. \ref{sec:results_and_discussion}, we show the results for the target systems obtained via the SFT and CCSD calculations.   
In Sect. \ref{sec:conclusions}, our conclusions are provided.

\section{Methods}
\label{sec:methods} 

\subsection{Second-quantized Hamiltonian}

For a practical SFT calculation of a realistic many-electron system,
we begin with the non-relativistic Hamiltonian for an isolated system in the first-quantized form given by
$H = H_{\mathrm{one}} +	H_{\mathrm{two}}$,
where
$
H_{\mathrm{one}}
=
-\sum_{i}
\nabla_i^2/2
-\sum_{i, \mu}
Z_\mu/ | \boldsymbol{r}_i - \boldsymbol{\tau}_\mu |
$
and
$
H_{\mathrm{two}}
=
\frac{1}{2}
\sum_{i, i'}
1/ | \boldsymbol{r}_i - \boldsymbol{r}_{i'} |
,
$
are the one- and two-body parts, respectively.
$\boldsymbol{r}_i$ is the position of the $i$th electron and $\boldsymbol{\tau}_\mu$ is the position of the atom $\mu$ with its atomic number $Z_\mu$.
We expand the one-electron spatial wave functions in localized basis functions at the individual atoms.
The basis functions are non-orthogonal in general.

In prior to the SFT calculation for the target system,
we perform a restricted Hartree--Fock (RHF) calculation to obtain orthonormalized spatial one-electron orbitals $\{ \phi_j \}$.
The subset of the HF orbitals near the Fermi level are adopted to prepare the correlated subspace, or equivalently the active space, for the subsequent SFT calculation.
Thus we need to construct the Hamiltonian in the second-quantized form
\begin{gather}
	\hat{H}
	=
		\sum_{\sigma, \sigma'}
		\sum_{j_1, j_2}
			t_{j_1 \sigma, j_2 \sigma'}
			\hat{a}_{j_1 \sigma}^\dagger
			\hat{a}_{j_2 \sigma'}
	\nonumber \\
		+
		\sum_{\sigma, \sigma'}
		\sum_{j_1, j_2, j_3, j_4}
			\frac{U_{j_1 j_2 j_3 j_4}}{2}
			\hat{a}_{j_1 \sigma}^\dagger
			\hat{a}_{j_2 \sigma'}^\dagger
			\hat{a}_{j_4 \sigma'}
			\hat{a}_{j_3 \sigma}
	.
	\label{H_2nd_for_HF_basis}
\end{gather}
The creation $\hat{a}_{j \sigma}^\dagger$ and the annihilation $\hat{a}_{j \sigma}$ operators of electrons for the HF orbitals satisfy the ordinary anticommutation relations because of their orthonormality.
The transfer integrals, or equivalently the one-electron integrals,
in eq. (\ref{H_2nd_for_HF_basis}) are the matrix elements $\langle \phi_{j_1} | H_{\mathrm{one}} | \phi_{j_2} \rangle$ of the one-body part between the HF orbitals.
These quantities are diagonal in spin space since the relativistic effects are not incorporated in the present study.
The Coulomb integrals, or equivalently the two-electron integrals,
in eq. (\ref{H_2nd_for_HF_basis}) are given by
\begin{gather}
	U_{j_1 j_2 j_3 j_4}
    =
		( \phi_{j_1} \phi_{j_3} | \phi_{j_2} \phi_{j_4} )
    \nonumber \\
	\equiv
		\int d^3 r d^3 r'
			\frac{
				\phi_{j_1} (\boldsymbol{r})
				\phi_{j_3} (\boldsymbol{r})
				\phi_{j_2} (\boldsymbol{r}')
				\phi_{j_4} (\boldsymbol{r}')
			}{ | \boldsymbol{r} - \boldsymbol{r}' | }
	.
\end{gather}
We have assumed the HF orbitals to be real.

Since the reference states for our CCSD calculations are the unrestricted Hartree--Fock (UHF) states,
the one- and two-electron integrals in the second-quantized Hamiltonian are spin dependent.

\subsection{CCSD calculations}

The CC state for a reference state $| \Psi_0 \rangle$ is constructed by performing an exponentially parametrized transform as $| \Psi_{\mathrm{CC}} \rangle \ = e^{\hat{T}} | \Psi_0 \rangle$,
where $\hat{T}$ is a so-called cluster operator.
The normalization of the CC wave function is difficult but is possible by adopting the bi-variational formulation,\cite{Arponen83,Bi-vari1,Bi-vari2}
with which we calculate the CCSD one-particle GFs\cite{Nooijen92,Nooijen93,Nooijen95} in the recently proposed procedure.\cite{Kowalski14,Kowalski16}
We use the UHF states as the reference states for CCSD calculations of the isolated atoms in the present study.

\subsection{SFT calculations}

\subsubsection{One-particle GF}

The one-particle temperature GF\cite{fetter2003quantum}, or equivalently the Matsubara GF, for interacting electrons is defined as
\begin{gather}
	G_{j \sigma, j \sigma'} (\tau)
    =
		-\langle \mathrm{T} \hat{a}_{j \sigma} (\tau) \hat{a}_{j' \sigma'}^\dagger (0) \rangle
        ,
	\label{def_temp_Green}
\end{gather}
where $\mathrm{T}$ is the time-ordering symbol for an imaginary time $-\beta < \tau < \beta$.
$\beta$ is the inverse temperature.
$
\hat{a}_{j \sigma}^\dagger (\tau)
\equiv
e^{\tau \hat{\mathcal{H}} } \hat{a}_{j \sigma}^\dagger e^{-\tau \hat{\mathcal{H}} }
$
and
$
\hat{a}_{j \sigma} (\tau)
\equiv
e^{\tau \hat{\mathcal{H}} } \hat{a}_{j \sigma} e^{-\tau \hat{\mathcal{H}} }
$
are the creation and annihilation operators, respectively,
in the Heisenberg's picture for the grand canonical Hamiltonian $\hat{\mathcal{H}} \equiv \hat{H} - \mu \hat{N}$.
$\mu$ is the chemical potential and $\hat{N}$ is the electron number operator.
Since the temperature GF is an anti-periodic function with period of $\beta$,
it is expanded in a Fourier series for the fermionic discrete Matsubara frequencies $\omega_n \equiv (2 n + 1) \pi/\beta$:
\begin{gather}
	G_{j \sigma, j' \sigma'} (\tau)
	=
		\frac{1}{\beta}
		\sum_{n = -\infty}^\infty
		e^{-i \omega_n \tau}
		G_{j \sigma, j' \sigma'} (i \omega_n)
        .
	\label{G_imag_time_freq_sum}
\end{gather}
The one-particle GF $G (z)$ for an arbitrary complex frequency $z$, however, can be defined by using analytic continuation for the spectral representation,
from which it is clear that
\begin{gather}
	G_{j \sigma, j' \sigma'} (z)
    =
		G_{j' \sigma', j \sigma} (z^*)^*
        .
	\label{Gz_conj_transpose}
\end{gather}

\subsubsection{Potthoff functional}

The grand potential of an interacting electronic system is expressed as the Baym--Kadanoff functional\cite{bib:3699},
which contains the Luttinger--Ward functional\cite{bib:3508} $\Phi^\mathrm{LW} [G]$.
By using the fact that the self-energy $\Sigma [G]$ is obtained via the functional derivative of the LW functional,
Potthoff\cite{bib:3694,Potthoff_VCA} proved that the grand potential is expressed as
\begin{gather}
	\Omega^\mathrm{P} [\Sigma]
    =
    	F [\Sigma]
        -
        \mathrm{Tr}
        \ln
        [ - G_0^{-1} + \Sigma]
        ,
   	\label{def_Potthoff_func}
\end{gather}
where $F [\Sigma]$ is the Legendre transform of the LW functional and $G_0$ is the non-interacting GF.
$\mathrm{Tr}$
is a shorthand notation of
$
\beta^{-1}
\sum_{n = -\infty}^\infty
e^{+i \omega_n 0}
\mathrm{tr}
,
$
where the trace is for the electronic states.
He demonstrated in addition that the necessary and sufficient condition for $\delta \Omega^\mathrm{P} [\Sigma] /\delta \Sigma$ to vanish is the Dyson equation,
\begin{gather}
	G^{-1}
    =
    	G_0^{-1}
        -
        \Sigma
	,
\end{gather}
that is,
the Potthoff functional gives a stationary value for the exact (physical) self-energy.
This variational principle allows one to introduce unphysical transfer integrals as variational parameters and/or fictitious one-particle orbitals for finding good trial self-energies.
We do not, however, introduce such parameters in the present study for simplicity of implementation.
It is noted that the stationarity condition for a self-energy functional does not ensure the convexity, that is, the physical self-energies can be saddle points.\cite{bib:3617}

Although the explicit functional form of $\Phi^{\mathrm{LW}} [G]$ is unknown and so is $F [\Sigma]$,
we can calculate the exact value of the unknown functional for the correlated subsystem from the results of exact diagonalization as
\begin{gather}
	F [\Sigma^{\mathrm{sub}}]
    =
    	\Omega^{\mathrm{sub}}
        +
        \mathrm{Tr}
        \ln
        [ - G_0^{\mathrm{sub} -1} + \Sigma^\mathrm{sub}]
        ,
	\label{F_for_corr_subs}      
\end{gather}
for which the grand potential is obtained from the partition function $Z^{\mathrm{sub}} = \exp (-\beta \Omega^{\mathrm{sub}} )$ and the self-energy $\Sigma^\mathrm{sub}$ from the Dyson equation for the subsystem.

In the present study,
we construct the self-energy $\Sigma$ for the whole system by embedding the subspace self-energy $\Sigma^{\mathrm{sub}}$ into the original Hilbert space block diagonally.
We assume that the one-particle states outside the subspace are completely filled or empty and hence the corresponding components in $\Sigma$ vanish.
For this assumption, it is reasonable to regard the value of the unknown functional for the whole system to be equal to that for the subsystem:
\begin{gather}
	F [\Sigma]
    =
		F [\Sigma^{\mathrm{sub}}]
        .
\end{gather}
We can then calculate the value of the Potthoff functional for the whole system by substituting the self-energy constructed in this way into eq. (\ref{def_Potthoff_func}).

\subsubsection{Decoupling of correlated subspace}

Due to large computational cost,
only a subset of the HF orbitals can be taken into account for exact diagonalization.
The correlated subspace thus has to be constructed by decoupling the adopted orbitals from the whole system.

The Coulomb integrals which involve the HF orbitals residing outside the correlated subspace have nonzero values in general.
They prevent us from solving the many-body problem for the subspace with the second-quantized Hamiltonian in its original form.
Therefore we adopt a mean-field approximation similar to that used  by Aichhorn et al.\cite{bib:3543}
Specifically, for a two-body term in eq. (\ref{H_2nd_for_HF_basis}) involving orbital(s) outside the subspace, 
we replace with four one-body terms as
\begin{gather}
	\frac{U_{j_1 j_2 j_3 j_4}}{2}
	\hat{a}_{j_1 \sigma}^\dagger
	\hat{a}_{j_2 \sigma'}^\dagger
	\hat{a}_{j_4 \sigma'}
	\hat{a}_{j_3 \sigma}
	\to
	\nonumber \\
		\frac{U_{j_1 j_2 j_3 j_4}}{2}
        \Bigg[
		\langle 
			\hat{a}_{j_2 \sigma'}^\dagger
			\hat{a}_{j_4 \sigma'}
		\rangle
		\hat{a}_{j_1 \sigma}^\dagger
		\hat{a}_{j_3 \sigma}
		+
        \langle 
			\hat{a}_{j_1 \sigma}^\dagger
			\hat{a}_{j_3 \sigma}
		\rangle
		\hat{a}_{j_2 \sigma'}^\dagger
		\hat{a}_{j_4 \sigma'}
	\nonumber \\
		-
        \langle 
			\hat{a}_{j_2 \sigma'}^\dagger
			\hat{a}_{j_3 \sigma}
		\rangle
		\hat{a}_{j_1 \sigma}^\dagger
		\hat{a}_{j_4 \sigma'}
		-
        \langle 
			\hat{a}_{j_1 \sigma}^\dagger
			\hat{a}_{j_4 \sigma'}
		\rangle
		\hat{a}_{j_2 \sigma'}^\dagger
		\hat{a}_{j_3 \sigma}
        \Bigg]
		-
		E^{\mathrm{dc}}_{j_1 j_2 j_3 j_4, \sigma \sigma'}
      	,
        \label{replacement_MFA}
\end{gather}
where
\begin{gather}
	E^{\mathrm{dc}}_{j_1 j_2 j_3 j_4, \sigma \sigma'}
	\equiv
		\frac{U_{j_1 j_2 j_3 j_4}}{2}
        \Bigg[
		\langle 
			\hat{a}_{j_1 \sigma}^\dagger
			\hat{a}_{j_3 \sigma}
		\rangle
		\langle 
			\hat{a}_{j_2 \sigma'}^\dagger
			\hat{a}_{j_4 \sigma'}
		 \rangle
         \nonumber \\
		-
        \langle 
			\hat{a}_{j_1 \sigma}^\dagger
			\hat{a}_{j_4 \sigma'}
		 \rangle
		\langle 
			\hat{a}_{j_2 \sigma'}^\dagger
			\hat{a}_{j_3 \sigma}
		\rangle
        \Bigg]
	\label{E_dc_from_each_term}        
\end{gather}
is the double-counting energy.
There exist two reasons for adopting this replacement.
First, the off-diagonal terms in spin space are necessary when studying the effects of electronic correlation on non-collinear magnetism.\cite{bib:3541, bib:3542, bib:3522}
Second, if we perform this kind of replacements for all the two-body terms,
the Hamiltonian will contain only the one-body terms and the electronic-structure calculation will reduce to an ordinary UHF calculation in quantum chemistry.
For a case where the one-particle orbitals are not HF ones, such a complete mean-field approximation does not necessarily reduce to an HF calculation. 

It is clear that the mean-field parameters, which are nothing but the one-particle density matrix,
are calculated from eqs. (\ref{def_temp_Green}) and (\ref{G_imag_time_freq_sum}) as
\begin{gather}
	\langle \hat{a}_{j \sigma}^\dagger \hat{a}_{j' \sigma'} \rangle
	=
		G_{j' \sigma', j \sigma} (\tau = -0)
	\nonumber \\        
	=
		\frac{1}{\beta}
		\sum_{n = -\infty}^\infty
		e^{+i \omega_n 0}
		G_{j' \sigma', j \sigma} (i \omega_n)
        .
	\label{mfp_as_Matsubara_sum}
\end{gather}
The expectation value of the electron number is calculated similarly as
$\langle \hat{N} \rangle = \mathrm{Tr} G (\tau = -0)$.
The summation for the infinite number of Matsubara frequencies in eq. (\ref{mfp_as_Matsubara_sum}) can be calculated as an integral on a complex contour plus summation for a finite number of Matsubara frequencies, as explained later.

The one-body Hamiltonian $\hat{H}_{\mathrm{one}}$ can be decomposed into the part $\hat{H}_{\mathrm{one}}^{\mathrm{sub}}$ involving only the operators for the orbitals in the correlated subspace and the residual part $\hat{H}_{\mathrm{one}}^{\mathrm{res}}$ as
\begin{gather}
	\hat{H}_{\mathrm{one}}
    =
	\hat{H}_{\mathrm{one}}^{\mathrm{sub}}
	+
	\hat{H}_{\mathrm{one}}^{\mathrm{res}}
    .
\end{gather}
It is similarly the case with $\hat{H}_{\mathrm{two}}$,
for which the replacements in eq. (\ref{replacement_MFA}) lead to the following one- and two-body parts:
\begin{gather}
	\hat{H}_{\mathrm{two}}
	=
		\hat{H}_{\mathrm{two}}^{\mathrm{sub}}
        +
		\hat{H}_{\mathrm{two}}^{\mathrm{res}}
	\to
		\hat{H}_{\mathrm{two}}^{\mathrm{sub}}
        +
		\hat{\widetilde{H}}_{\mathrm{one}}^{\mathrm{sub}}
        +
		\hat{\widetilde{H}}_{\mathrm{one}}^{\mathrm{res}}
        ,
\end{gather}
where we exclude the double-counting energies.
$\hat{\widetilde{H}}_{\mathrm{one}}^{\mathrm{sub}}$ consists only of the one-body operators for the orbitals in the correlated subspace
and $\hat{\widetilde{H}}_{\mathrm{one}}^{\mathrm{res}}$ is the residual part.
By using them, we construct the mean-field-approximated Hamiltonian of the correlated subspace
\begin{gather}
	\hat{H}_{(\mathrm{MF})}^{\mathrm{sub}}
    \equiv
		\hat{H}_{\mathrm{one}}^{\mathrm{sub}}
        +
		\hat{\widetilde{H}}_{\mathrm{one}}^{\mathrm{sub}}
        +
		\hat{H}_{\mathrm{two}}^{\mathrm{sub}}
\end{gather}
and that of the whole system
\begin{gather}
	\hat{H}_{(\mathrm{MF})}
    \equiv
		\hat{H}_{\mathrm{one}}^\mathrm{res}
        +
		\hat{\widetilde{H}}_{\mathrm{one}}^\mathrm{res}
        +
		\hat{H}_{(\mathrm{MF})}^\mathrm{sub}
        \label{def_whole_H_opr_MF}
	.        
\end{gather}
We adopt the self-energy calculated from $\hat{H}_{(\mathrm{MF})}^{\mathrm{sub}}$ as the subspace self-energy:
\begin{gather}
	\Sigma^{\mathrm{sub}}
    \equiv
		\Sigma_{(\mathrm{MF})}^{\mathrm{sub}}
        ,
	\label{Sigma_MF_as_Sigma}      
\end{gather}
to be substituted into eq. (\ref{F_for_corr_subs}).
The grand potential in the mean-field approximation is thus calculated from eq. (\ref{def_Potthoff_func}) as
\begin{gather}
	\Omega^\mathrm{P}_{(\mathrm{MF})} [\Sigma]
    =
		\Omega^\mathrm{P} [\Sigma]
    	-
        E^{\mathrm{dc}}
        .
\end{gather}
$E^{\mathrm{dc}}$ is the total double-counting energy calculated by summing up all the contributions in eq. (\ref{E_dc_from_each_term}).
We emphasize here that there can exist alternatives other than eq. (\ref{Sigma_MF_as_Sigma}) for obtaining better trial self-energies for the whole system.
It will be interesting to look for more elaborate construction of the subspace Hamiltonian by employing various many-body theories.

We denote $\hat{H}_{(\mathrm{MF})}^\mathrm{sub}$ by $\hat{H}^\mathrm{sub}$ for simplicity in what follows.

\subsubsection{Green's function of correlated subspace}

By choosing the correlated subspace comprised of a moderate number of one-particle orbitals near the Fermi level in order for exact diagonalization to be achievable,
we can construct the exact one-particle GF for the subspace.
Its expression at a finite temperature is written as\cite{bib:3066} 
\begin{gather}
	G_{m m'}^\mathrm{sub} (z)
	=
		\frac{1}{Z^{\mathrm{sub}}}
		\sum_\nu
			e^{-\beta ( \varepsilon_\nu - \mu N_\nu)}
			[
				G_{m m'}^{(\mathrm{e}) \nu} (z)
				+
				G_{m m'}^{(\mathrm{h}) \nu} (z)
			]
            ,
	\label{G_imag_freq_sum_partial_G}
\end{gather}
where
\begin{gather}
	G_{m m'}^{(\mathrm{e}) \nu} (z)
	=
		\langle \nu | \hat{a}_m 
		\frac{1}{ z + \mu + \varepsilon_\nu - \hat{H}^\mathrm{sub}}
		\hat{a}_{m'}^\dagger | \nu \rangle
	\label{def_partial_G_e}
\end{gather}
and
\begin{gather}
	G_{m m'}^{(\mathrm{h}) \nu} (z)
	=
		\langle \nu | \hat{a}_{m'}^\dagger 
		\frac{1}{ z + \mu - \varepsilon_\nu + \hat{H}^\mathrm{sub}}
		\hat{a}_m | \nu \rangle
	\label{def_partial_G_h}
\end{gather}
are the partial GFs for electron and hole excitations, respectively,
from the $\nu$th energy eigenstate containing $N_\nu$ electrons in the subspace.
$m$ and $m'$ are the composite indices of the one-electron orbitals and spins spanning the subspace.
The summation on the right-hand side in eq. (\ref{G_imag_freq_sum_partial_G}) are for all the energy eigenstates for all the possible electron numbers.
Since the expression eq. (\ref{G_imag_freq_sum_partial_G}) is given as the sum of the contributions containing the Boltzmann factors,
it is expected that only few number of energy eigenstates with relatively low eigenvalues
needs to be taken into account for an accurate calculation of the GF.
The smallest number of the involved eigenstates for a given accuracy, however,
depends on the temperature and/or the other parameters for the system being considered, as pointed out by Capone et al.\cite{bib:3066}
In a practical calculation,
we thus need to introduce a threshold $\varepsilon_{\mathrm{B}}$ for the Boltzmann factors to truncate the summation in eq. (\ref{G_imag_freq_sum_partial_G}),
for which we check the convergence of the results with respect to the threshold carefully.

The individual components of the GF can be calculated by employing the Lanczos method\cite{bib:3064} by using auxiliary quantities similarly to our previous study.\cite{GF_CCSD_and_FCI_for_light_atoms}

It is noted that the basic ideas stated above are much common to the complete active space (CAS)\cite{bib:4456} and its related approaches.
The various improvements for CAS proposed so far will help to sophisticate the treatment of correlated subspaces in SFT calculations.

\subsubsection{Complex-contour integrals}
\label{complex_integ_for_Matsubara}

\begin{figure}
\includegraphics[width=8cm]{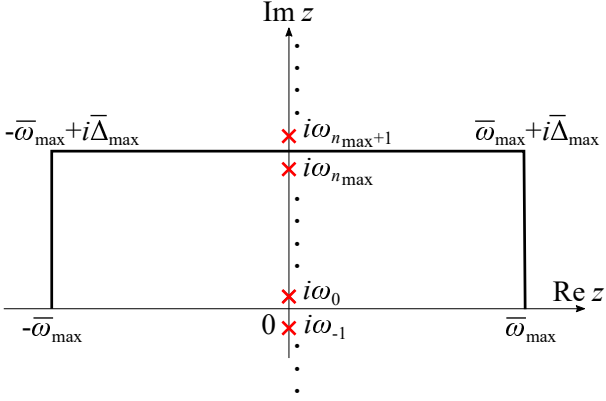}
%\vspace{40mm}
\caption{Complex-integration paths as thick lines and the Matsubara frequencies as crosses for an efficient calculation of the Matsubara sum in eq. (\ref{integ_for_Matsubara_sum}).}
\label{Fig_complex_integ}
\end{figure}

For the calculations of the sums over the infinite number of Matsubara frequencies [see eq. (\ref{mfp_as_Matsubara_sum})],
we adopt the approach proposed by Lu and Arrigoni.\cite{bib:3501}
For an arbitrary complex function $g (z)$ of a complex variable $z$
which satisfies that
$g(z) \to 0$ for $|z| \to \infty$ and $g (z^*) = g (z)^*$,
their approach enables one to calculate the Matsubara sum
$
I
\equiv
\beta^{-1}
\sum_{n = -\infty}^{\infty}
e^{+i \omega_n 0}
g (i \omega_n)
$
by using the numerical integration on a complex contour and the simple summation over a finite number of Matsubara frequencies.
For the off-diagonal components of the Green's function used in the present work, however,
$G_{m m'} (z^*) \ne G_{m m'} (z)^*$ in general.
We therefore define 
\begin{gather}
	G_{m m'}^{(+)} (z) \equiv G_{m m'} (z) + G_{m' m} (z)
	\\
	G_{m m'}^{(-)} (z) \equiv i G_{m m'} (z) - i G_{m' m} (z)
    ,
\end{gather}
which clearly satisfy
$G_{m m'}^{(\pm)} (z^*) = G_{m m'}^{(\pm)} (z)^*$ due to eq. (\ref{Gz_conj_transpose}).
The formula proposed by Lu and Arrigoni can then be applied to the sums $I_{m m'}^{(\pm)}$ for $G_{m m'}^{(\pm)} (z)$.
Specifically, the deformation of the complex contour enables one to rewrite the sum for the function $g$ to\cite{bib:3501}
\begin{gather}
	I
	=
		- \frac{1}{\pi}
		\int_{-\overline{\omega}_{\mathrm{max}}}^{\overline{\omega}_{\mathrm{max}}}
			d \omega
			\mathrm{Im} F (\omega + i \overline{\Delta}_{\mathrm{max}})
	\nonumber \\            
		- \frac{1}{\pi}
		\int_0^{\overline{\Delta}_{\mathrm{max}}}
			dy
			\mathrm{Re} F (-\overline{\omega}_{\mathrm{max}} + i y)
	\nonumber \\            
		+ \frac{1}{\pi}
		\int_0^{\overline{\Delta}_{\mathrm{max}}}
			dy
			\mathrm{Re} F ( \overline{\omega}_{\mathrm{max}} + i y)
		+ \frac{2}{\beta}
		\sum_{n = 0}^{n_{\mathrm{max}}}
			\mathrm{Re} g (i \omega_n)
	,
    \label{integ_for_Matsubara_sum}
\end{gather}
where $F (z) \equiv g(z)/(e^{\beta z} + 1)$ is the product of $g$ and the complex Fermi distribution function.
$\overline{\omega}_{\mathrm{max}}$ is an arbitrary positive constant satisfying
$\mathrm{Im} g (\omega + i 0) = 0$
for $|\omega| > \overline{\omega}_{\mathrm{max}}$.
$\overline{\Delta}_{\mathrm{max}}$ is an arbitrary positive constant satisfying
$\omega_{n_{\mathrm{max}}} < \overline{\Delta}_{\mathrm{max}} < \omega_{n_{\mathrm{max}} + 1}$
for the largest index $n_{\mathrm{max}}$ of the Matsubara frequency chosen for numerical tractability, as shown in Fig. \ref{Fig_complex_integ}.  
By using the Matsubara sums calculated in this way, we can obtain the sums for $G_{m m'} (z)$ and $G_{m' m} (z)$ as
\begin{gather}
	I_{m m'}
	=
		\frac{1}{2} [  I_{m m'}^{(+)} - i I_{m m'}^{(-)} ]
	\\
	I_{m' m}
	=
		\frac{1}{2} [  I_{m m'}^{(+)} + i I_{m m'}^{(-)} ]
	.
\end{gather}
We use this prescription also for the Matsubara sums for the correlated subspace.

The prescription described above could also be applied to other many-body methods such as GF2\cite{bib:4432,bib:3578},
which involves the Matsubara sums in the Galitskii--Migdal formula.\cite{stefanucci2013nonequilibrium}

\subsubsection{Calculation procedure}

\begin{figure}
\includegraphics[width=7cm]{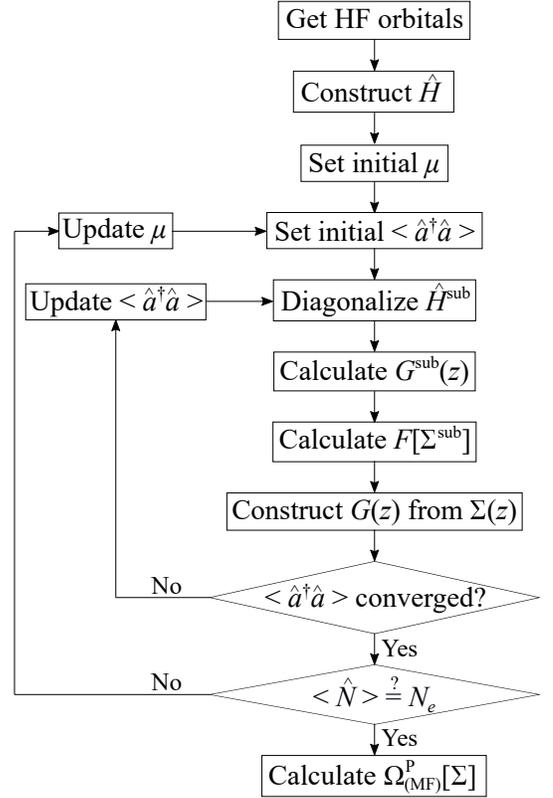}
%\vspace{40mm}
\caption{Procedure of a self-consistent SFT calculation in the present study.}
\label{Fig_sft_procedure}
\end{figure}

Since our Hamiltonian $\hat{H}_{(\mathrm{MF})}$ in eq. (\ref{def_whole_H_opr_MF}) depends on the mean-field parameters,
which are calculated from the GF [see eq. (\ref{mfp_as_Matsubara_sum})],
those parameters have to be determined self-consistently for a given chemical potential.
Furthermore, the chemical potential has to be determined so that the expected electron number is equal to the true electron number $N_e$ of the target system.
The problem thus has to be solved self-consistently undergoing the double loops.
The procedure of a self-consistent SFT calculation is shown in Fig. \ref{Fig_sft_procedure}.

\section{Computational details}
\label{sec:computational_details} 

We adopt the 6-31G basis set\cite{bib:4369} for the Cartesian Gaussian-type basis functions\cite{Helgaker} of all the elements in the present study.
The number of basis functions is thus 29 per spin direction for each atom.
The Coulomb integrals between them are calculated efficiently.\cite{Libint1}
By transforming them using the results of the RHF calculations,
we obtain the integrals between the HF orbitals.

In the SFT calculations, we perform exact diagonalization of the Hamiltonian matrices for correlated subspaces by using the Arnoldi method.\cite{bib:ARPACK}
For rapid convergence of the mean-field parameters,
we adopt the modified Broyden's method, proposed by Johnson.\cite{bib:3145}
This method is known to accelerate the convergence for correlated electronic systems.\cite{bib:3143}

\section{Results and discussion}
\label{sec:results_and_discussion}

\subsection{Total energies}

By allowing for fractional occupancy of each HF orbital,
we first performed RHF calculations for isolated
V  ($3 d^3 4 s^2$),
Cr ($3 d^5 4 s^1$),
Mn ($3 d^5 4 s^2$),
Fe ($3 d^6 4 s^2$),
Co ($3 d^7 4 s^2$),
Ni ($3 d^8 4 s^2$), and
Cu ($3 d^{10} 4 s^1$) atoms
with the imposed electronic configurations in the parentheses obtained in previous numerical HF calculations.\cite{bib:4368}
We obtained the RHF solutions which realize the spherically symmetric self-consistent electron densities,
from which we constructed the second-quantized Hamiltonians for the subsequent SFT calculations.
On the other hand, we performed UHF calculations by imposing integer occupancies to get the reference states for the CCSD calculations.
The total energies for the HF calculations are shown in Table \ref{table_energies}.

\begin{table*}
\centering
\caption{Total energies (a.u.) of isolated atoms for spherical RHF,  ordinary UHF, SFT from the RHF states, and CCSD from the UHF states. $S$ represents the spin for the UHF calculations. $E_{\mathrm{gs}}$ is the expected energy for $n_{\mathrm{gs}}$-fold degenerate ground states at 1000K calculated from eq. (\ref{relation_gpot_and_energy}).}
\label{table_energies}
\begin{tabular}{lrlcrlccclcc} 
\hline
                       & \multicolumn{1}{c}{RHF} &  & \multicolumn{2}{c}{UHF}                           &  & \multicolumn{3}{c}{SFT from RHF}                                                               &  & \multicolumn{2}{c}{CCSD from UHF}                                                         \\ 
\cline{2-2}\cline{4-4}\cline{5-5}\cline{7-7}\cline{8-8}\cline{9-9}\cline{11-11}\cline{12-12}
                       & \multicolumn{1}{c}{$E$} &  & $S$                     & \multicolumn{1}{c}{$E$} &  & $n_{\mathrm{gs}}$                & correlated subspace                     & $E_{\mathrm{gs}}$ &  & correlated subspace               & $E$                                                   \\ 
\hline
\multicolumn{1}{c}{V}  & -942.0047               &  & 3/2                     & -942.7875               &  & 28                               & $3d$                                    & -942.7547         &  & full                              & -942.8974                                             \\
                       & \multicolumn{1}{l}{}    &  & \multicolumn{1}{l}{}    & \multicolumn{1}{l}{}    &  & \multicolumn{1}{l}{}             & \multicolumn{1}{l}{$3s + 3p + 3d + 4s$} & -942.8030         &  & $3s + 3p + 3d + 4s$               & -942.8141                                             \\
\multicolumn{1}{c}{Cr} & -1042.0606              &  & 3                       & -1043.1922              &  & 7                                & $3d + 4s$                               & -1043.1628        &  & full                              & \begin{tabular}[c]{@{}c@{}}-1043.2547\\\end{tabular}  \\
                       & \multicolumn{1}{l}{}    &  & \multicolumn{1}{l}{}    & \multicolumn{1}{l}{}    &  & \multicolumn{1}{l}{}             & $3s + 3p + 3d + 4s$                     & -1043.1814        &  & $3s + 3p + 3d + 4s$               & \begin{tabular}[c]{@{}c@{}}-1043.1922\\\end{tabular}  \\
\multicolumn{1}{c}{Mn} & -1148.4943              &  & 5/2                     & -1149.7221              &  & 6                                & $3d$                                    & -1149.6865        &  & full                              & -1149.7964                                            \\
                       & \multicolumn{1}{l}{}    &  & \multicolumn{1}{l}{}    & \multicolumn{1}{l}{}    &  & \multicolumn{1}{l}{}             & $3s + 3p + 3d + 4s$                     & -1149.7081        &  & $3s + 3p + 3d + 4s$               & -1149.7221                                            \\
\multicolumn{1}{c}{Fe} & -1261.0986              &  & 2                       & -1262.2670              &  & 25                               & $3d$                                    & -1262.2417        &  & full                              & -1262.3737                                            \\
                       & \multicolumn{1}{l}{}    &  & \multicolumn{1}{l}{}    & \multicolumn{1}{l}{}    &  & \multicolumn{1}{l}{}             & $3s + 3p + 3d + 4s$                     & -1262.2552        &  & $3s + 3p + 3d + 4s$               & -1262.2670                                            \\
\multicolumn{1}{c}{Co} & -1380.1388              &  & 3/2                     & -1381.1978              &  & 28                               & $3d$                                    & -1381.1831        &  & full                              & -1381.3172                                            \\
                       & \multicolumn{1}{l}{}    &  & \multicolumn{1}{l}{}    & \multicolumn{1}{l}{}    &  & \multicolumn{1}{l}{}             & $3s + 3p + 3d + 4s$                     & -1381.1879        &  & $3s + 3p + 3d + 4s$               & -1381.1978                                            \\
\multicolumn{1}{c}{Ni} & -1505.7926              &  & 1                       & -1506.6095              &  & 21                               & $3d$                                    & -1506.6027        &  & full                              & \begin{tabular}[c]{@{}c@{}}-1506.7560\\\end{tabular}  \\
                       & \multicolumn{1}{l}{}    &  & \multicolumn{1}{l}{}    & \multicolumn{1}{l}{}    &  & \multicolumn{1}{l}{}             & $3s + 3p + 3d + 4s$                     & -1506.6027        &  & $3s + 3p + 3d + 4s$               & -1506.6096                                            \\
\multicolumn{1}{c}{Cu} & -1638.5044              &  & 1/2                     & -1638.5670              &  & 2                                & $3d + 4s$                               & -1638.5649        &  & full                              & -1638.7535                                            \\
                       & \multicolumn{1}{l}{}    &  & \multicolumn{1}{l}{}    & \multicolumn{1}{l}{}    &  & \multicolumn{1}{l}{}             & $3s + 3p + 3d + 4s$                     & -1638.5649        &  & $3s + 3p + 3d + 4s$               & -1638.5670                                            \\
\hline
\end{tabular}
\end{table*}

We performed the SFT calculations for the second-quantized Hamiltonian in the representation of the spherical RHF orbitals, in contrast to the CCSD calculation with the UHF reference states.
It is because that SFT has been developed for describing the statistically averaged electronic properties and the isolated atoms are in spherically symmetric environments.
We confirmed that the chemical potential $\mu$ set to zero for our SFT calculations gives the correct expected electron numbers for all the target systems.
We used a temperature of 1000K for numerical tractability.
The ground states for each isolated atom are degenerate as an $LS$ multiplet.
The partition function incorporating only the $n_{\mathrm{gs}}$-degenerate ground states with the energy eigenvalue $E_{\mathrm{gs}}$ is $Z = n_{\mathrm{gs}} \exp(- \beta E_{\mathrm{gs}})$.
The grand potential and the expected energy for the zero chemical potential are thus connected via the relation
\begin{gather}
	\Omega = E_{\mathrm{gs}} - \frac{1}{\beta} \ln n_{\mathrm{gs}}
    ,
    \label{relation_gpot_and_energy}
\end{gather}
where the second term on the right-hand side is nothing but the entropy term.
The expected SFT energies are shown in Table \ref{table_energies} together with the CCSD total energies.
We found that the correct number of grounds states of the correlated subspace for each atom was obtained in the exact diagonalization.
The SFT expected energy of each atom using the $3 s + 3 p + 3 d + 4s$ subspace is lower than that using the $3 d (+ 4 s)$ subspace, as expected.
The full-CCSD total energy for each atom is the lowest since all the combinations of orbitals over different subspaces are allowed.
It is interesting to notice that the SFT and CCSD energies for the $3 s + 3 p + 3 d + 4s$ subspaces are close to each other for the individual systems despite the quite different formalism of the two approaches.

\subsection{Orbital energies}

\begin{figure}
\includegraphics[width=8.5cm]{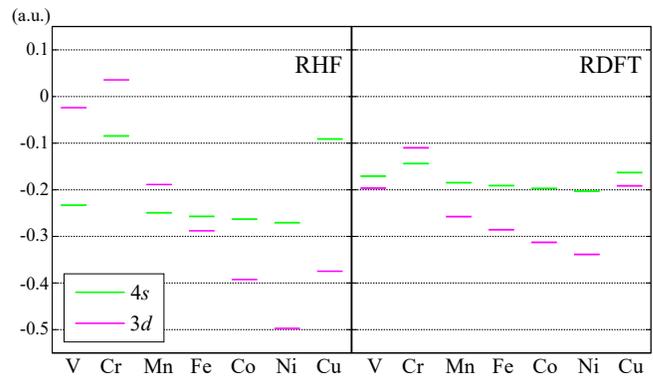}
%\vspace{40mm}
\caption{
Orbital energies of $3 d$ and $4 s$ orbitals obtained in (a) spherical RHF and (b) spherical RDFT-GGA calculations by allowing for fractional occupancies.
Electron configurations used for the RDFT calculations are the same as for the RHF calculations.
}
\label{Fig_orb_ene_3d4s}
\end{figure}

Before moving to the analysis of many-body spectra,
let us take look at the one-particle orbital energies.
The orbital energies of $3 d$ and $4 s$ orbitals obtained in the RHF calculations are shown in Fig. \ref{Fig_orb_ene_3d4s}.
We also show those obtained in spherical RDFT calculations for comparison,
in which the radial KS equations were solved exactly w	ith the GGA functional\cite{PBE96} with the same electron configurations as the RHF calculations.

The aufbau principle, that require electrons occupy the energy levels from lower to higher, is not satisfied for all the atoms other than the V, Mn, and Cu atoms for the RHF orbitals, as seen in the figure.
It is also the case for all the atoms other than the Cu atom for the RDFT orbitals.
These observations tell us that how the one-particle spectra for an isolated atom look like is not trivial even for such a simple system.
It is thus necessary to calculate the spectra for the atoms by using the sophisticated methods incorporating the correlation effects to elucidate the shapes and origins of their spectral structures.

\subsection{Spectral functions}

The spectral function of an electronic system is given as the imaginary part of the GF on real axis in frequency domain:
\begin{gather}
	A (\omega)
    =
    	-
        \frac{1}{\pi}
        \mathrm{Im} [\mathrm{tr} G (\omega + i \delta)]
	,
\end{gather}
where $\delta$ is an infinitesimal positive constant for ensuring causality, set to 0.01 a.u. in our calculations.
The major peaks below and above the Fermi level are called the quasiparticle peaks, corresponding to the removal and addition of an electron to a target system, respectively.

\begin{figure*}
\includegraphics[width=16cm]{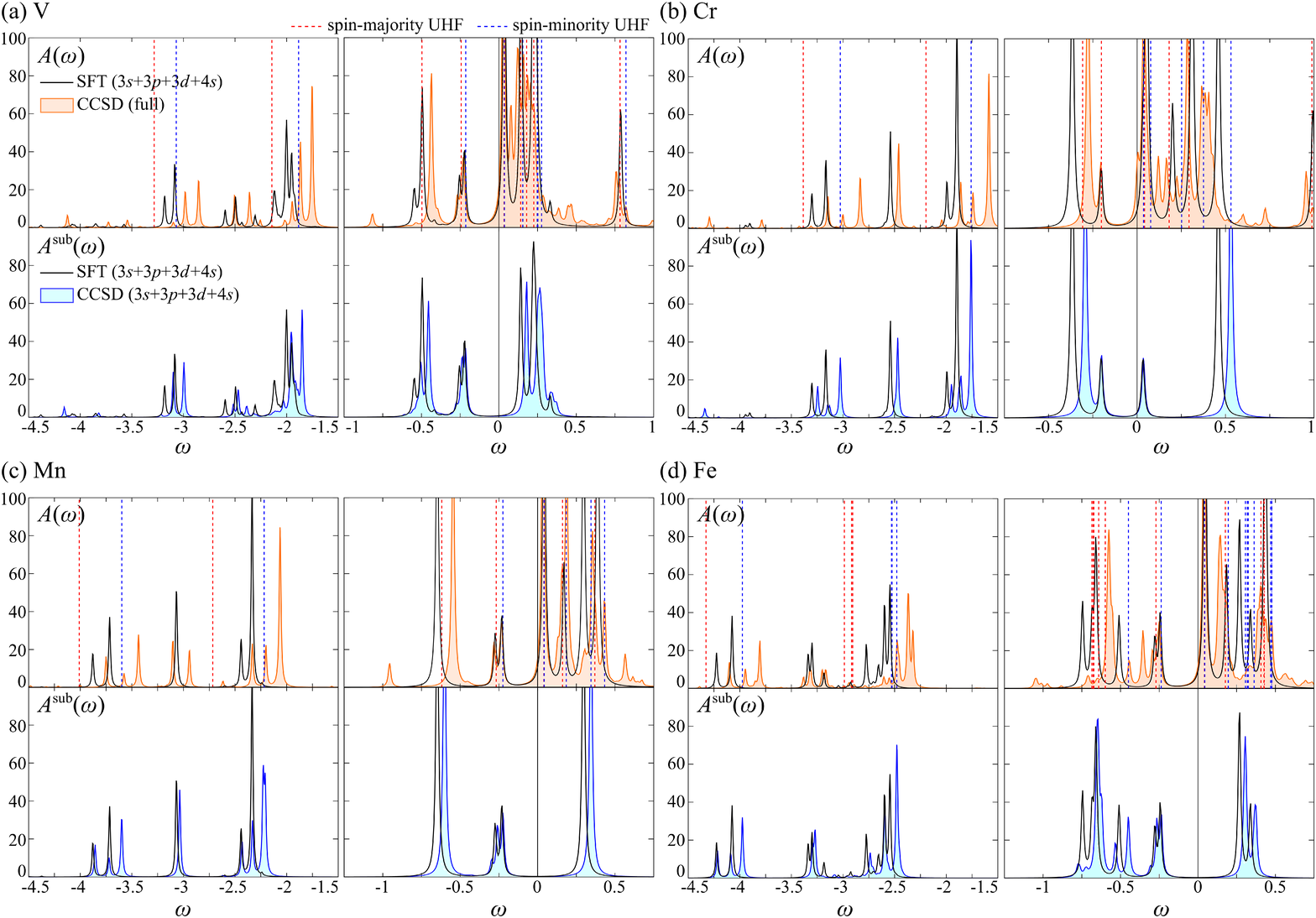}
%\vspace{40mm}
\caption{
Spectral functions $A (\omega)$ (in a.u.) 
as functions of $\omega$ (in a.u.)
for isolated V, Cr, Mn, and Fe atoms calculated from the SFT calculations using the $3 s + 3 p + 3 d + 4 s$ correlated subspaces and those from the full CCSD calculations. 
Spectral functions $A^{\mathrm{sub}} (\omega)$ for the correlated subspaces from the SFT calculations and those from the CCSD calculations restricted within the subspaces are also shown.
Vertical dashed lines represent the UHF orbital energies.
}
\label{Fig_spectra_V_Cr_Mn_Fe}
\end{figure*}

\begin{figure*}
\includegraphics[width=16cm]{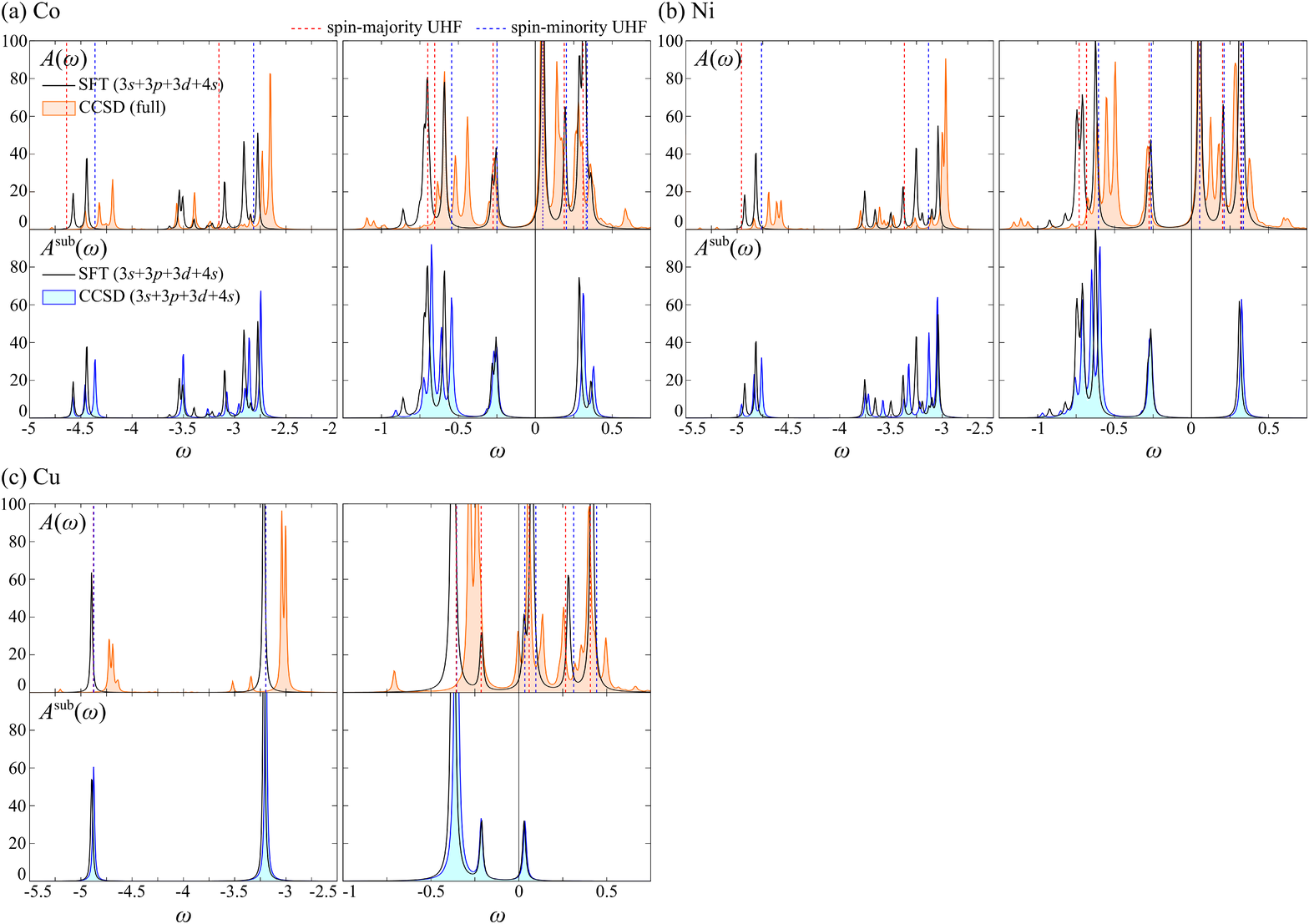}
%\vspace{40mm}
\caption{
Spectral functions for isolated Co, Ni, Cu atoms similar to Fig. \ref{Fig_spectra_V_Cr_Mn_Fe}.
}
\label{Fig_spectra_Co_Ni_Cu}
\end{figure*}

The spectral functions for the atoms calculated from the SFT calculations using the $3 s + 3 p + 3 d + 4 s$ correlated subspaces and those from the full CCSD calculations are shown in Figs. \ref{Fig_spectra_V_Cr_Mn_Fe} and \ref{Fig_spectra_Co_Ni_Cu}.
The spectral functions $A^{\mathrm{sub}} (\omega)$ for the correlated subspaces from the SFT calculations and those from the CCSD calculations restricted within the subspaces are also shown.
The spectra far above the Fermi level is not very reliable since the localized basis functions we are using cannot describe the free-electron-like continuum.
We therefore focus on the analyses of the spectra below the Fermi level.

The assignment of spectral peaks in $A (\omega)$ for each atom is described in Appendix \ref{assignment_spec}.
We have found that the quasiparticle peaks of occupied $3 d$ orbitals are \textcolor{red}{always} below those of occupied $4 s$ peaks for each atom, in contrast to the RHF spectra for V, Cr, and Mn atoms and the RDFT spectra for Cr atom (see Fig. \ref{Fig_orb_ene_3d4s}). 

The SFT spectra for the whole system exhibit the $3 s$ and $4 s$ quasiparticle peaks which can be easily assigned to the UHF orbital energies with exchange splittings,
though the SFT calculations started from the RHF states.
It indicates that the self-consistency loop of SFT for each atom has reached correctly the degenerate ground states, each of which is spin polarized.
On the other hand, the spectra near $3 p$ states exhibit many satellite peaks, looking more complicated than the UHF spectra, spread in wider energy ranges.
The number of satellite peaks for each quasiparticle peak is larger in the full CCSD calculations than in the SFT calculations.
It is because that the larger number of one-particle orbitals for more excitation channels were took into account in the full CCSD than in the correlated subspace for the SFT calculations,
consistent with the picture for light atoms examined by us.\cite{GF_CCSD_and_FCI_for_light_atoms}

The $3 d$ and $4 s$ spectra, which are found near the Fermi level for each atom,
for the Cr, Mn, and Cu atoms have simpler forms than the other atoms.
They are understood by remembering that the spin-majority $3 d$ orbitals in Cr and Mn atoms are fully occupied and the spin-minority ones are empty,
realizing the spherical electron densities.
The electron density in Cu atom is spherical as well since the $3 d$ orbitals for both spin directions are fully occupied.
The spherical shapes of these three atoms make their $3 d$ spectra look simple,
while those of the other atoms exhibit complicated structures coming from the anisotropic electron density of each of the degenerate ground states.
The exchange splittings in Cu atom are quite smaller than the other atoms since only the $4 s$ orbitals are partially occupied.
These results indicate that the spin-polarized and/or anisotropic ground states have to be taken into account in an SFT calculation from an RHF state even for a spherically symmetric system for avoiding the breaking of aufbau principle in Fig. \ref{Fig_orb_ene_3d4s}.
On the other hand, an UHF state corresponding to any one of the degenerate ground states has to be used as the reference state for the CCSD calculation.
The SFT calculations are thus physically appropriate from this viewpoint compared to the CCSD calculations,
that is,
the former do not require any reference state with artificially broken symmetry and the results retain the spherical symmetry via the statistical average over the multiple symmetry-broken ground states.

We can find the overall peak structures of the subspace spectra obtained by the SFT and CCSD calculations look quite similar to each other.
It is because the CCSD GFs for the given subspaces can be regarded as those obtained via insufficient exact diagonalization for the subspaces.
We can find, however, differences between the subspace spectra, which are the peak shift of the occupied states in the CCSD subspace spectra to higher frequencies than that of the SFT spectra,
implying that the self-consistency loop for CCSD has led to more significant effects for narrowing gaps than that for SFT calculations with our mean-field approximation.

\begin{figure}
\includegraphics[width=4.5cm]{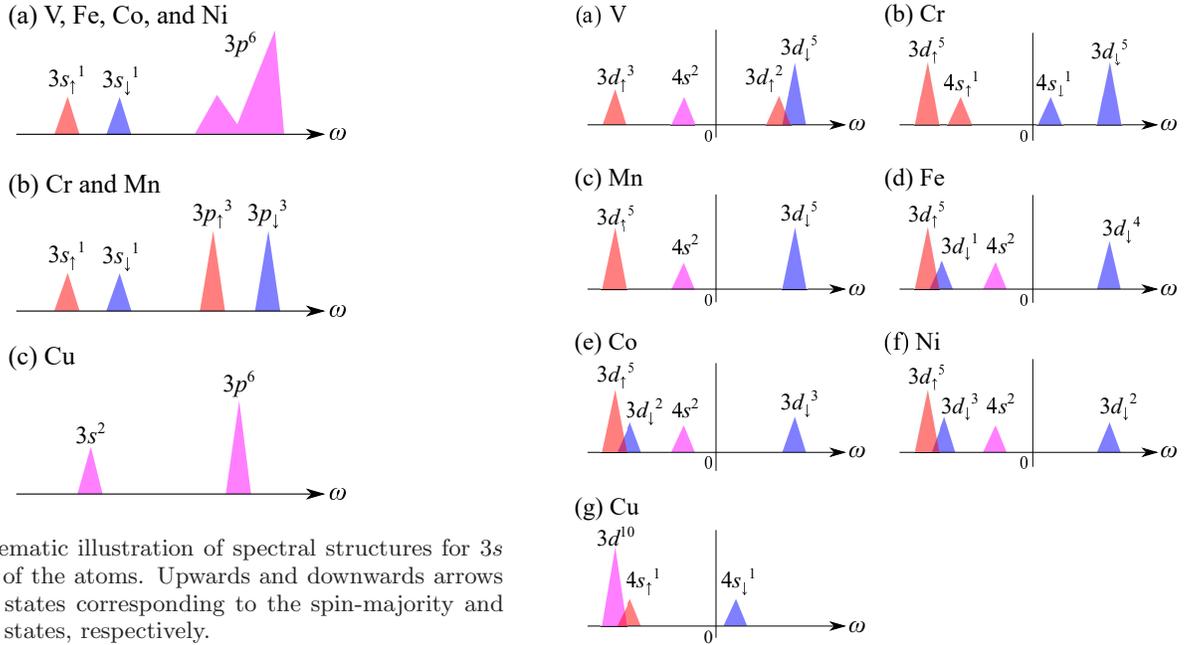}
%\vspace{40mm}
\caption{
Schematic illustration of spectral structures for $3 s$ and $3 p$ peaks of the atoms.
Upwards and downwards arrows represent the states corresponding to the spin-majority and spin-minority states, respectively.  
}
\label{Fig_schematic_3s3p}
\end{figure}

\begin{figure}
\includegraphics[width=8cm]{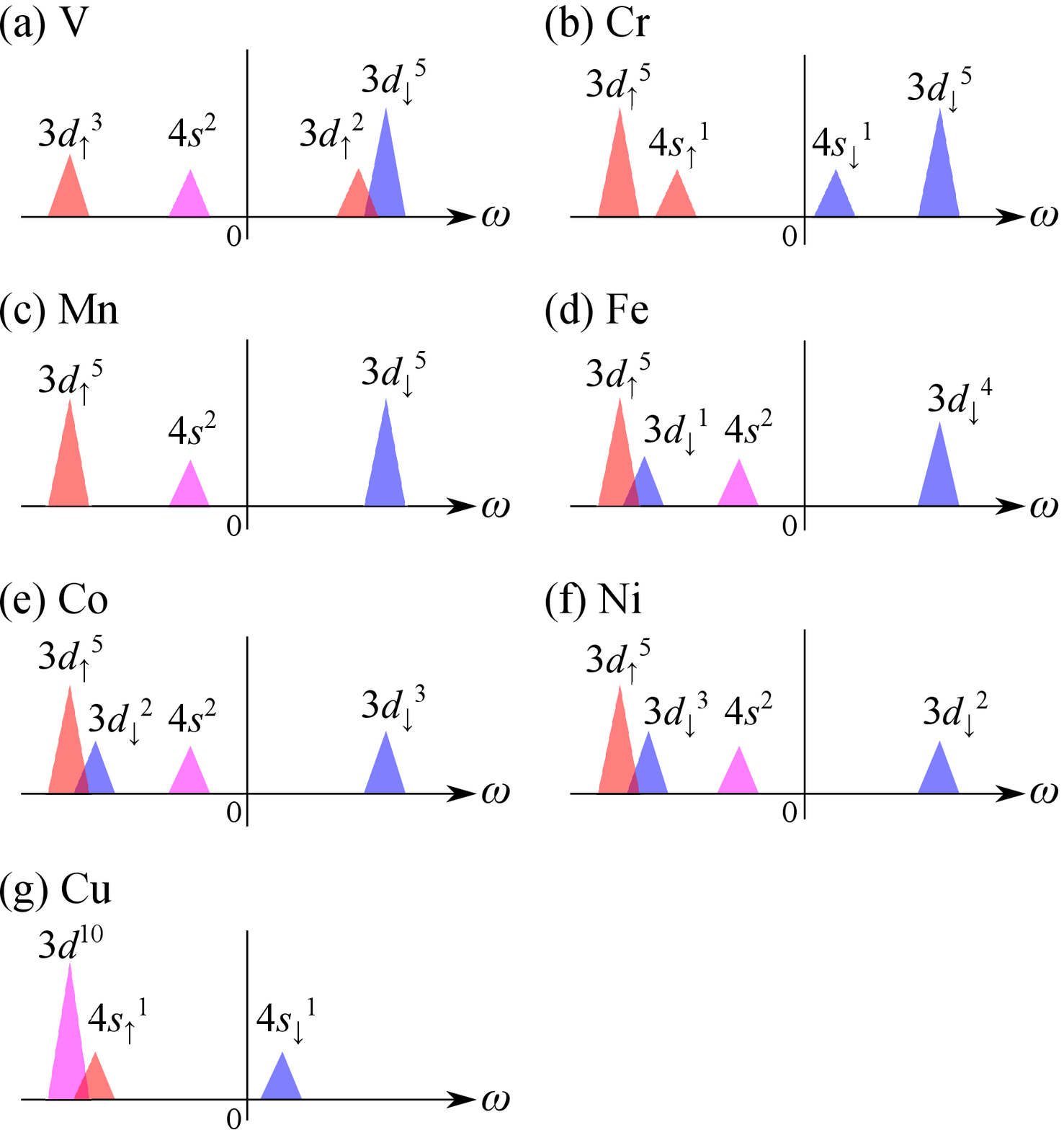}
%\vspace{40mm}
\caption{
Schematic illustration of spectral structures for $3 d$ and $4 s$ peaks of the atoms similar to Fig. \ref{Fig_schematic_3s3p}.
}
\label{Fig_schematic_3d4s}
\end{figure}

The spectral structures observed in our many-body calculations are schematically illustrated in Figs. \ref{Fig_schematic_3s3p} and \ref{Fig_schematic_3d4s}.
Although these understandings are similar to those which can be extracted from the UHF picture,
we can find some deviation of the full-CCSD spectra from such rough UHF picture.
In fact, for the Co and Ni atoms, the spin-majority and spin-minority $3d$ quasiparticle peaks and the significant satellite peaks are close to each other below Fermi level to form the complicated spectra.
In addition, for the Cu atom, the $3d$ peaks and the spin-majority $4s$ ones are quite close to each other than in the UHF spectra.

\section{Conclusions}
\label{sec:conclusions}

We demonstrated that self-consistent calculations based on SFT are possible for the electronic structure of realistic systems.
We described the procedure of an SFT calculation in detail.

We performed the calculations for isolated $3 d$ transition metal atoms from V to Cu as a preliminary study.
We compared the total energies and one-particle GFs obtained in this way and those obtained from the CCSD method. 
The full-CCSD total energy for each atom was found to be the lowest.
The SFT and CCSD energies for the same correlated subspaces were close to each other for the individual systems despite the quite different formalism of the two approaches.

We found not only the quasiparticle peaks but also satellite peaks in the one-particle SFT and CCSD GFs as the many-body effects.
The exchange splittings correctly appeared in the SFT GFs despite the RHF initial states.
The CCSD GFs exhibited more complicated satellite structures than the SFT GFs since the former incorporated more excitation channels.
By referring to the RHF and RDFT orbital energies of the isolated atoms,
we demonstrated for the correct spectral features,
that the spin-polarized and/or anisotropic ground states have to be taken into account even for a spherically symmetric system.
The SFT calculations correctly retained the spherical symmetry of the atoms via the statistical average over the multiple symmetry-broken ground states in contrast to the CCSD calculations.

The mean-field approximation adopted in the present study is only one of possible alternatives for the construction of a correlated subspace.
There can exist many other sophisticated ways for improving the correlated subspace and the resultant self-energy for the whole system such as making use of the variational property of the Potthoff functional and the unitary rotations over different subspaces.
Particularly, various improvements and examinations for the SEET approach will give us useful insights since the SFT approach can be regarded as a kind of it.
Since the basic ideas for correlated subspaces are much common to CAS and its related approaches,
the various improvements for CAS proposed so far will help to sophisticate the treatment of correlated subspaces in SFT calculations.

\begin{acknowledgments}
This research was supported by MEXT as ``Exploratory Challenge on Post-K computer" (Frontiers of Basic Science: Challenging the Limits). This research used computational resources of the K computer provided by the RIKEN Advanced Institute for Computational Science through the HPCI System Research project (Project ID: hp170261). Y.M. acknowledges the support from JSPS Grant-in-Aid for Young Scientists (B) (Grant No. 16K18075).
\end{acknowledgments}

\appendix

\section{Regularization of divergence}

The second term on the right-hand side of eq. (\ref{def_Potthoff_func}),
that of eq. (\ref{F_for_corr_subs}),
and the mean-field parameters in eq. (\ref{mfp_as_Matsubara_sum}) have to be calculated.
Since $G(z) \propto 1/z$ for a large $z$, however,
the Matsubara sum involving the GF does not converge if we adopt the as-is expression.
Thus we have to regularize the divergence by introducing the divergence regularization Green's function\cite{Potthoff_VCA}
\begin{gather}
	G_{\mathrm{reg} m m'}(z)
	\equiv
		\delta_{m m'}
		\frac{1}{z}
	,
\end{gather}
which is formally equal to the Green's function of a non-interacting system whose energy eigenvalues are all equal to the chemical potential.
Since the Matsubara sum for an non-interacting system is equal to the grand potential,
it can be calculated as\cite{fetter2003quantum}
\begin{gather}
	S_{\mathrm{reg}}
	\equiv
		-\frac{1}{\beta}
		\sum_{n = -\infty}^{\infty}
			e^{+ i \omega_n 0}
			\ln \det [ -G_{\mathrm{reg}}^{-1} (i \omega_n) ]
	\nonumber \\
	=
		-\frac{2}{\beta}
		n_{\mathrm{orbs}}
		\ln 2
	,
	\label{Matsubara_sum_G_reg}
\end{gather}
where the factor $2$ is the spin degeneracy and $n_{\mathrm{orbs}}$ is the number of spatial one-electron orbitals.
The appropriate divergence-regularized expression of the Matsubara sum for eq. (\ref{def_Potthoff_func}) thus reads
\begin{gather}
	S
	=
		S_{\mathrm{reg}}
		-\frac{1}{\beta}
		\sum_{n = -\infty}^{\infty}
			e^{+ i \omega_n 0}
			(
				\ln \det [ -G^{-1} (i \omega_n) ]
	\nonumber \\
				-
				\ln \det [ -G_{\mathrm{reg}}^{-1} (i \omega_n) ]
			)
	.
	\label{Matsubara_sum_det_G_reglarized}
\end{gather}
We adopt this expression in the present study since the summation on the right-hand side above converges.
We use this prescription also for the Matsubara sums in eqs. (\ref{F_for_corr_subs}) and (\ref{mfp_as_Matsubara_sum}).

\section{Assignment of spectral peaks}
\label{assignment_spec}

We assign the significant peaks below the Fermi level for the SFT and CCSD spectra in
Figs. \ref{Fig_spectra_V_Cr_Mn_Fe} and \ref{Fig_spectra_Co_Ni_Cu}
to the quasiparticle ones or the associated satellite ones by comparing the UHF orbital energies as follows.
For the V atom,
the $3 s$ peaks for $-3.25 < \omega < -2.75$ a.u.,
the $3 p$ peaks for $-2.75 < \omega < -1.5$ a.u.,
the $3 d$ peaks for $-1 < \omega < -0.375$ a.u., and
the $4 s$ peaks for $-0.375 < \omega < 0$ a.u.
For the Cr atom,
the $3 s$ peaks for $-3.5 < \omega < -2.75$ a.u.,
the $3 p$ peaks for $-2.75 < \omega < -1.5$ a.u.,
the $3 d$ peaks for $-0.5 < \omega < -0.25$ a.u., and
the $3 d$ peaks for $-0.25 < \omega < 0$ a.u.
For the Mn atom,
the $3 s$ peaks for $-4 < \omega < -3.25$ a.u.,
the $3 p$ peaks for $-3.25 < \omega < -2$ a.u.,
the $3 d$ peaks for $-1 < \omega < -0.375$ a.u., and
the $4 s$ peaks for $-0.375 < \omega < 0$ a.u.
For the Fe atom,
the $3 s$ peaks for $-4.5 < \omega < -3.5$ a.u.,
the $3 p$ peaks for $-3.5 < \omega < -2$ a.u., and
the $3 d$ and $4 s$ peaks for $-1.25 < \omega < 0$ a.u.
For the Co atom,
the $3 s$ peaks for $-4.75 < \omega < -4$ a.u.,
the $3 p$ peaks for $-3.75 < \omega < -2.5$ a.u.,
the $3 d$ peaks for $-1.25 < \omega < -0.375$ a.u., and
the $4 s$ peaks for $-0.375 < \omega < 0$ a.u.
For the Ni atom,
the $3 s$ peaks for $-5.25 < \omega < -4.5$ a.u.,
the $3 p$ peaks for $-4 < \omega < -2.5$ a.u.,
the $3 d$ peaks for $-1.25 < \omega < -0.375$ a.u., and
the $4 s$ peaks for $-0.375 < \omega < 0$ a.u.
For the Cu atom,
the $3 s$ peaks for $-5.25 < \omega < -4.5$ a.u.,
the $3 p$ peaks for $-4 < \omega < -2.5$ a.u., and
the $3 d$ and $4 s$ peaks for $-1 < \omega < 0$ a.u.

\bibliographystyle{apsrev4-1}
\bibliography{paper}

\end{document}